\documentclass[fleqn]{2023SCGE}

\usepackage[numbers, sort&compress]{natbib}
\begin{document}

\ensubject{subject}

\ArticleType{Article}
\SpecialTopic{SPECIAL TOPIC: }
\Year{2026}
\Month{March}
\Vol{66}
\No{1}
\DOI{??}
\ArtNo{000000}
\ReceiveDate{March xx, 2026}
\AcceptDate{xx x, 2026}

\title{Direct Detection of Type~II-P Supernova Progenitors with the \textit{Euclid} and CSST Surveys}{Direct Detection of Type~II-P Supernova Progenitors with the \textit{Euclid} and CSST Surveys}

\author[1,2]{Junjie Wu}{}
\author[2,1,3]{Ning-Chen Sun}{sunnc@ucas.ac.cn}
\author[2,1]{Zexi Niu}{}
\author[1,2]{Tianmang Zhang}{}
\author[4,5,6]{Chun Chen}{}
\author[7,1]{Xiaohan Chen}{}
\author[8,9]{\\Nancy Elias-Rosa}{}
\author[10]{Morgan Fraser}{}
\author[2,1]{Xinyi Hong}{}
\author[11]{Justyn Maund}{}
\author[2,1]{César Rojas-Bravo}{}
\author[2,1]{Anyu Wang}{}
\author[1]{\\Beichuan Wang}{}
\author[2,1]{Ziyang Wang}{}
\author[2,1]{Qiang Xi}{}
\author[2,1]{Linxi Zhang}{}
\author[1,2]{Yinuo Zhang}{}

\AuthorMark{Wu J J}

\AuthorCitation{Wu J J, Sun N C, Niu Z X, et al}

\address[1]{National Astronomical Observatories, Chinese Academy of Sciences, Beijing, 100101, China}
\address[2]{School of Astronomy and Space Science, University of Chinese Academy of Sciences, Beijing, 100049, China}
\address[3]{Institute for Frontiers in Astronomy and Astrophysics, Beijing Normal University, Beijing, 102206, China}
\address[4]{School of Physics and Astronomy, Sun Yat-sen University, Zhuhai, 519082, China}
\address[5]{CSST Science Center for the Guangdong-Hong Kong-Macau Greater Bay Area, Sun Yat-sen University, Zhuhai, 519082, China}
\address[6]{Dipartimento di Fisica, Universita` di Napoli ``Federico II'', Compl. Univ. di Monte S. Angelo, Via Cinthia, Napoli, I-80126, Italy}
\address[7]{School of Physics and Astronomy, China West Normal University, Nanchong, 637002, China}
\address[8]{INAF – Osservatorio Astronomico di Padova, vicolo dell’Osservatorio 5, Padova, 35122, Italy}
\address[9]{Institute of Space Sciences (ICE, CSIC), Campus UAB, Carrer de Can Magrans s/n, Barcelona, 08193, Spain}
\address[10]{School of Physics, University College Dublin, L.M.I. Main Building, Beech Hill Road, Dublin 4, D04 P7W1, Ireland}
\address[11]{Department of Physics, Royal Holloway, University of London, Egham, TW20 0EX, United Kingdom}

\abstract{Identifying and characterizing supernova (SN) progenitor stars remains a central yet difficult goal in SN research, limited by archival images lacking sufficient depth or spatial resolution and circumstellar dust biasing intrinsic parameter estimates. This field will be revolutionized by \textit{Euclid} and the upcoming Chinese Space-station Survey Telescope (CSST), which conduct deep, wide-field, high-resolution and multi-band imaging surveys. We evaluate their detection capability by comparing model magnitudes of RSG progenitors with detection limits, finding their optical and near-infrared filters highly effective. Monte-Carlo simulations predict that completed \textit{Euclid} and CSST surveys will enable $\lesssim$13 (or 24) progenitor detections per year within the mass range of 8--16 (or 8--25)\,$M_\odot$, an order of magnitude higher than the current detection rate of $\sim$1 per year (primarily based on HST). With the circumstellar dust, the emerging spectral energy distribution (SED) of the SN progenitor is mainly affected by the optical depth and is almost independent of dust temperature in their survey filters. Mock tests demonstrate that the progenitor mass and dust optical depth can be derived simultaneously by fitting the observed SED over 11 survey filters while fixing dust temperature to a typical value. \textit{Euclid} and CSST will significantly enlarge the sample of direct progenitor detections with accurate mass measurements, crucial for resolving the long-standing RSG problem.}



\maketitle

\begin{multicols}{2}
\section{Introduction}
Core-collapse supernovae (CCSNe), the explosive deaths of massive stars, play a vital role in shaping galaxies through chemical enrichment, feedback into the interstellar medium, and the formation of compact remnants such as neutron stars and black holes. They also serve as critical laboratories for testing theories of massive star evolution and as multi-messenger sources of neutrinos, gravitational waves, and cosmic rays \cite{Warren-2020}. It is a central goal in SN research to identify and characterize SN progenitors. Direct progenitor detection by comparing pre-explosion images with post-explosion observations has so far yielded only a handful of confirmed cases, primarily from archival \textit{Hubble} Space Telescope (HST) data (e.g., \cite{Kilpatrick-2017eaw-2018, Kilpatrick-2019yvr-2021, Kilpatrick-2023ixf-2023, Maund-2005, Maund-2011, Maund-2014, Niu-2023ixf-2023, Niu-2017gkk-2024, Niu-2024abfo-2025, Niu-2026, Smartt-2009, VanDyk-2013df-2014, VanDyk-2017eaw-2019, VanDyk-2023ixf-2024, Xiang-2017ein-2019, Xiang-2023ixf-2024, Xiang-2024ggi-2024}).

These observations have revealed red supergiant (RSG) progenitors for the majority of Type~II-P SNe, yet all measured initial masses fall below $\sim$16--18\,$M_\odot$, despite stellar evolution models predicting that RSGs can exist up to $\sim$25--30\,$M_\odot$ \cite{Smartt-2015}. This discrepancy, commonly referred to as the ``RSG problem'', remains unresolved in massive star evolution. It is possible that the most massive RSGs ($M\gtrsim18\,M_\odot$) do not produce typical Type~II-P SNe. Instead, they may explode as other types of SNe (e.g., Type~II-L, Type~IIn or Type~II-b) \cite{Leonard-2011, Groh-2013, Suzuki-2025} or may not yield a visible transient at all. In the latter case, these stars could undergo failed explosions, collapsing directly into black holes without a bright signal, due to insufficient energy in the neutrino-driven revival of the stalled shock \cite{Nadezhin-1980, Lovegrove-2013, Kashiyama-2015, Sukhbold-2016}. Therefore, study of the RSG problem is crucial to our understanding of the final fate of massive stars and their explosion physics.

The current sample of detected Type~II-P SN progenitors, however, remains extremely small (fewer than 30 secure identifications), raising concerns about the statistical significance of the RSG problem, although some work \cite{Rodriguez-2022} argues otherwise. With such limited numbers, the apparent upper mass cutoff could simply reflect sampling variance rather than a true physical boundary \cite{Davies-2018}. Moreover, circumstellar dust formed in pre-explosion mass-loss episodes can cause significant, wavelength-dependent extinction that is not accounted for in standard foreground-only extinction corrections \cite{Walmswell-2012}. If uncorrected, this extra dimming leads to underestimates of progenitor luminosities and, consequently, initial masses derived from evolutionary tracks.

Therefore, it is very important to build a large sample of direct progenitor detections with well-determined progenitor parameters. Currently, the scarcity of progenitor detections stems from stringent observational requirements: deep, high-resolution images of nearby galaxies before explosion. HST, while offering excellent resolution, covers only a tiny fraction of the sky, severely limiting the number of observed galaxies and thus the yield of detectable progenitors.
Moreover, progenitors identified in archival HST images are frequently detected in only one or two filters, providing insufficient spectral information to reliably constrain circumstellar extinction. In many cases, the circumstellar extinction is highly uncertain, and often implicitly assumed to be zero, despite significant dust formation often occurring in the dense winds of massive stars.

This field will be revolutionized by the advent of the \textit{Euclid} mission and the Chinese Space-station Survey Telescope (CSST). \textit{Euclid} is a medium-class mission in the European Space Agency’s (ESA) Cosmic Vision 2015--2025 programme, launched in July 2023. Operating from the Sun–Earth L2 Lagrange point, it carries a 1.2-meter telescope equipped with two main scientific instruments: the visible imaging instrument (VIS) and the Near Infrared Spectrometer and Photometer (NISP). VIS delivers high-resolution optical imaging in the $I$-band over a 0.54\,deg$^2$ field of view (FoV) with $0.1''\mathrm{pixel}^{-1}$ \cite{Euclid-vis-2025}, while NISP provides near-infrared (NIR) imaging in $Y$, $J$,\,$H$ bands over a 0.57\,deg$^2$ FoV with $0.3''\mathrm{pixel}^{-1}$ and slitless spectroscopy \cite{Euclid-nisp-2025}. Over its nominal 6-year mission, \textit{Euclid} will conduct the \textit{Euclid} Wide Survey (EWS), covering 14\,679\,deg$^2$ (or 13,245 \,deg$^2$ due to the optical stray light problem) of extragalactic sky with unprecedented depth ($I=26.3$,\,$Y=24.6$,\,$J=24.6$,\,$H=24.5$ AB mag for $5\sigma$ point source; \cite{Euclid-vis-2025, Euclid-nisp-2025}) and spatial resolution of 0.13$''$, 0.33$''$, 0.35$''$, and 0.35$''$ in the respective bands \cite{Euclid-overview-2025}.

Complementing \textit{Euclid} in the ultraviolet (UV) and optical domains, the upcoming CSST promises a similarly transformative advance. Scheduled for launch in 2027, CSST features an off-axis three-mirror anastigmat (TMA) design with a 2-meter primary mirror and carries five main scientific instruments. It will conduct a deep, multi-band imaging survey across seven photometric bands ($NUV=25.4$,\,$u=25.4$,\,$g=26.3$,\,$r=26.0$,\,$i=25.9$,\,$z=25.2$, and $y=24.4$ AB mag for $5\sigma$ point source) over a wide FoV of 1.1\,deg$^2$ with $0.074''\mathrm{pixel}^{-1}$, achieving a spatial resolution of $\sim0.20''$, ultimately covering up to 17\,500\,deg$^2$ of the sky over a 10-year life cycle \cite{CSST-2026, CSST-Optical-Ban-2026, CSST-MS-overview-Wei-2026}. CSST will operate in a co-orbit with the China Space Station (CSS) at an altitude of approximately 400\,km. This unique orbital configuration allows CSST to periodically rendezvous and dock with the CSS for on-orbit servicing (OOS), enabling instrument upgrades, maintenance, and repairs, thereby ensuring long-term operational flexibility and scientific longevity \cite{zhan-2011, zhanhu-2021}.

\begin{figure}[H]
    \centering
    \includegraphics[width=\columnwidth]{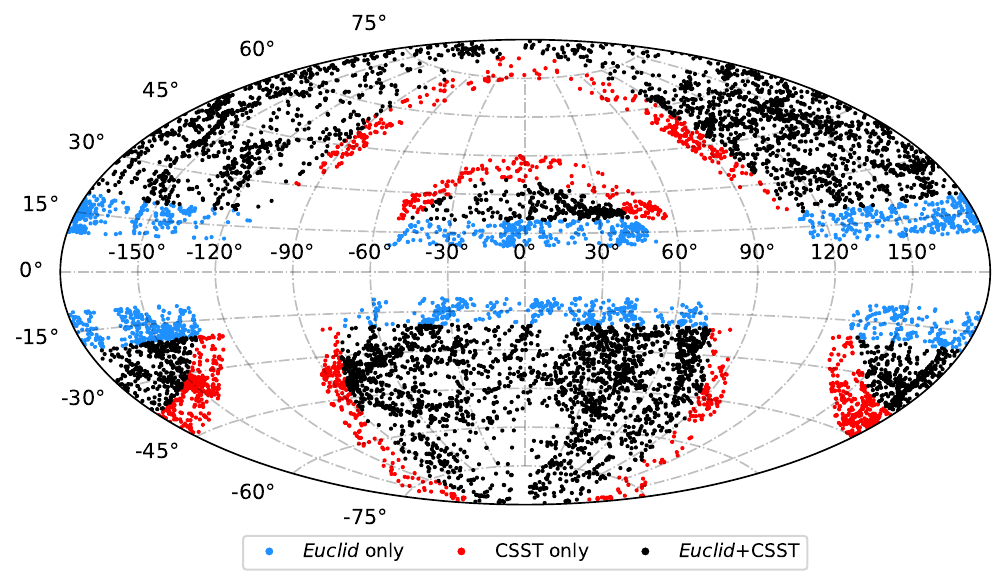}
    \caption{Local galaxies within the \textit{Euclid} and CSST survey footprints, shown in barycentric mean ecliptic coordinates. Galaxies within the \textit{Euclid}+CSST area are marked in black, those within the \textit{Euclid}-only area in blue, and those within the CSST-only area in red.}
    \label{fig:survey_area}
\end{figure}

\begin{figure*}[!ht]
    \centering
    \includegraphics[width=\textwidth]{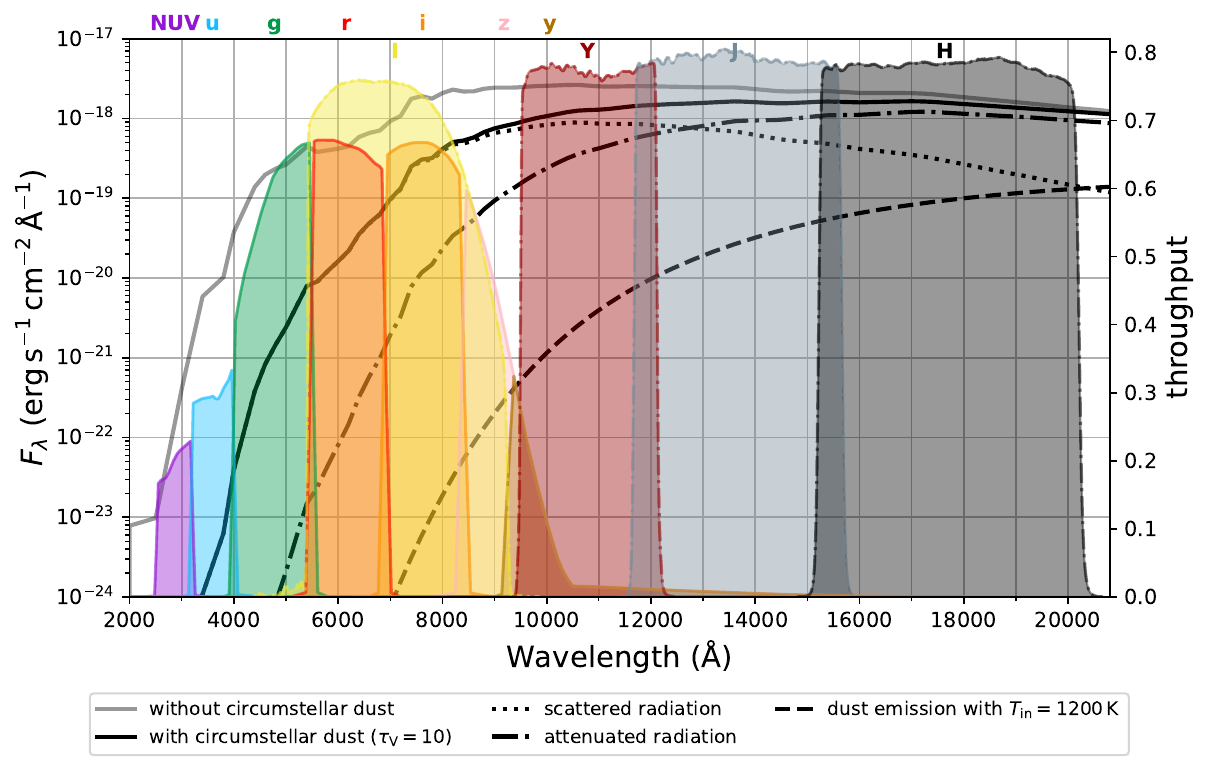}
    \caption{Sensitivity curves for \textit{Euclid} and CSST filters, compared with synthetic SEDs of an RSG ($M_\mathrm{ini}=15\,M_{\odot}$) embedded in a spherically symmetric oxygen-rich dust envelope ($\tau_\mathrm{V}=10$). Two different inner boundary temperature conditions are considered: $T_{\mathrm{in}}=200$\,K and 1200\,K. Colored shaded regions display the total sensitivity of photometric bands in the \textit{Euclid} and CSST surveys. Solid, dotted, dash-dotted, and dashed lines correspond to the the total emerging radiation, scattered radiation, attenuated radiation and the dust emission, respectively. Most radiation components exhibit minimal sensitivity to variations in $T_{\mathrm{in}}$, resulting in overlapping SED curves for different temperatures. In contrast, the dust emission shows significant dependence on $T_{\mathrm{in}}$. Notably, the dust emission for $T_{\mathrm{in}}=200\,$K falls well below the lower limit of the y-axis. The flux densities shown are obtained assuming the RSG is located at a distance of $d=10\,{\rm Mpc}$.}
    \label{fig:sed_m_sun_15}
\end{figure*}

While the primary science goals of \textit{Euclid} and CSST lie in cosmology, they also offer valuable opportunities for the identification of SN progenitors in the nearby galaxies of the local universe. These surveys will provide high-resolution images over a large sky area of 17\,500~deg$^2$. Based on this legacy dataset, SN progenitor searches will be possible for a much larger number of nearby galaxies than before (see Figure~\ref{fig:survey_area}). Moreover, the \textit{Euclid} and CSST will perform their surveys in a total of 11 filters in the UV, optical, and NIR bands, covering a very wide wavelength range from $\sim2000\,\text{Å}$ to $\sim20\,000\,\text{Å}$. This will tightly constrain the spectral energy distribution (SED) of the SN progenitors, allowing for accurate determination of their properties (see Figure~\ref{fig:sed_m_sun_15}).

In this work, we assess the capability of the \textit{Euclid} and CSST surveys in detecting Type~II-P SN progenitors and explore how to derive their fundamental parameters based on these detections. This paper is structured as follows. In Section~\ref{capability} we compare the brightness of SN progenitors with the detection limits of \textit{Euclid} and CSST to estimate the number of progenitors detections each year. Section~\ref{parameters} analyzes the influence of circumstellar extinction on the observed SED and presents a possible method to constrain progenitor mass in the presence of circumstellar extinction. This paper is finally closed with a summary and conclusions.

\section{Detection Capability for RSG Progenitors}\label{capability}

\subsection{Detection Limits}

\begin{figure*}[!ht]
    \centering
    \includegraphics[width=1.0\textwidth]{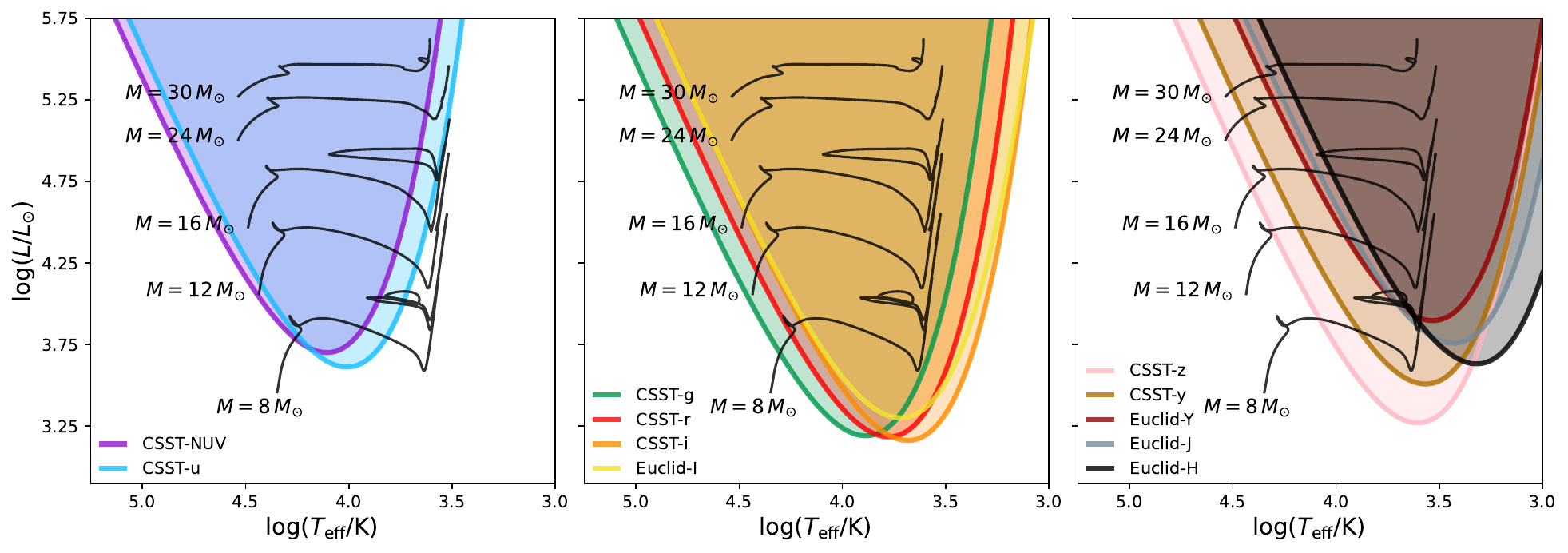}
    \caption{The evolutionary tracks of massive stars in comparison to the \textit{Euclid}/CSST detection limits. Black solid lines show the evolutionary tracks of massive stars with initial masses of 8, 12, 16, 24, and 30\,$M_\odot$ obtained from the \textsc{parsec} stellar evolution models. Colored solid curves indicate the typical detection limits for the \textit{Euclid} and CSST surveys at 10\,Mpc.}
    \label{fig:evo-hrd-bin}
\end{figure*}

\begin{figure*}[!ht]
    \centering
    \includegraphics[width=\textwidth]{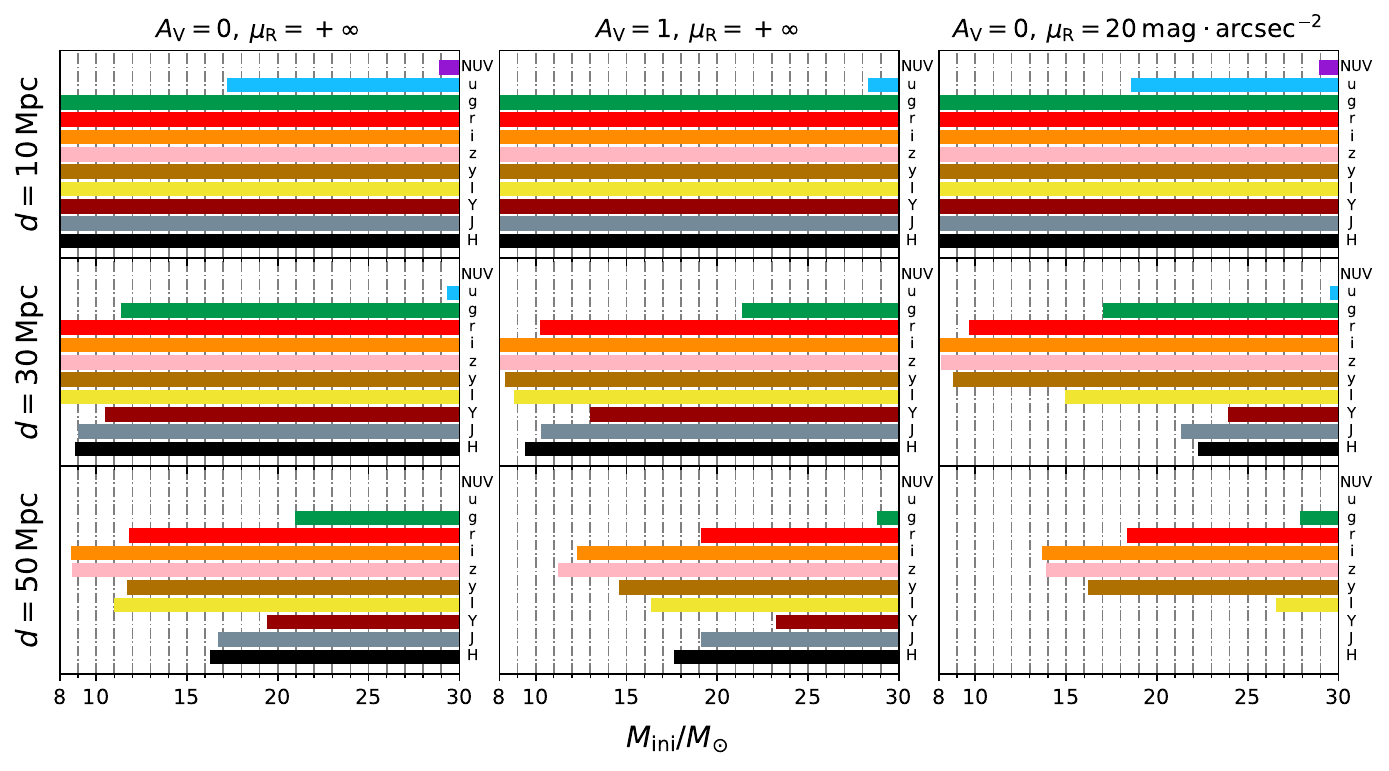}
    \caption{Mass range of SN progenitors detectable by \textit{Euclid} and CSST surveys. This figure is divided into nine panels. Three columns represent $A_{\mathrm{V}}=0$ with no host-galaxy background, $A_{\mathrm{V}}=1$ with no host-galaxy background, and $A_{\mathrm{V}}=0$ with a $R$-band background surface brightness of $20\,\mathrm{AB}\,\mathrm{mag\,arcsec^{-2}}$. Three rows correspond to distances of 10, 30, and 50\,Mpc, respectively. Each panel illustrates the detectable mass range of SN progenitors under the specified extinction and host-galaxy background conditions.}
    \label{fig:detect}
\end{figure*}

It is very important to assess the detection capability for CCSN progenitors by \textit{Euclid} and CSST surveys. For simplicity, we firstly consider CCSN progenitors at a distance of $d=10\,{\rm Mpc}$ without any extinction or host-galaxy background. Subsequently, we develop exposure time calculators for determining the signal-to-noise ratio (SNR) and limiting magnitude of point sources based on the photometric parameters of \textit{Euclid} and CSST surveys \cite{Euclid-overview-2025, CSST-2026}. Figure~\ref{fig:evo-hrd-bin} shows the \textsc{parsec} (V2.1; \cite{parsec-Bressan-2012, parsec-Chen-2014, parsec-Chen-2015, parsec-Fu-2018}) evolutionary tracks for solar-metallicity massive stars in comparison to the \textit{Euclid}/CSST 5$\sigma$ detection limits under such conditions. Here we have assumed a blackbody SED and calculate the synthetic magnitudes with the \textsc{pysynphot} package \cite{pysynphot-2013}. In the \textsc{parsec} models, massive stars up to 30\,$M_\odot$ evolve to RSGs before CCSNe, while those above 30\,$M_\odot$ become Wolf-Rayet (WR) stars. Some observational studies suggest that the turn-off mass required to produce WN stars (nitrogen-sequence WR stars) is at least 25\,$M_\odot$ in the Milky Way, rising to above 30\,$M_\odot$ in the Large Magellanic Cloud, and is likely closer to 35-40\,$M_\odot$ for the formation of WC stars (carbon-sequence WR stars) \cite{Massey-2001, Massey-2003, Crowther-2007, Smartt-2009}. It is worth noting that the precise boundary between RSG and WR stars can be critically affected by the highly uncertain mass-loss prescription in stellar models \cite{Woosley-2002, Heger-2003, Levesque-2005, Levesque-2006}.

Most of the RSG progenitors fall below the 5$\sigma$ detection limits of both the $NUV$ and $u$ filters of CSST; only the extremely massive RSG progenitors with $M_{\mathrm{ini}}\sim30\,M_\odot$ (or $\sim24\,M_\odot$) marginally exceed the $NUV$-band (or $u$-band) limit. This is because the RSG progenitors have very low effective temperatures (3,300-4,000\,K; \cite{Smartt-2015}) and most of their radiation lies at long wavelengths (see Figure~\ref{fig:sed_m_sun_15}). The other optical and NIR filters, on the contrary, are very useful in probing the cool RSGs. At $d=10$\,Mpc and without extinction or host-galaxy background, these filters can detect RSG progenitors even down to the lowest mass limit for CCSNe of $M_{\mathrm{ini}}=8\,M_\odot$.

Whether an RSG progenitor can be detected also depends on its distance and extinction, which influences its brightness in different filters, as well as the host-galaxy background, which affects the detection limits of \textit{Euclid}/CSST. To assess their influence, we repeat the above analysis and derive the lowest mass of RSG progenitors that can be detected under different conditions (see Figure~\ref{fig:detect}). In this context, a detection is defined as the identification of a source in at least one photometric band with SNR $\geq5$. We consider three different distances of $d=10,\,30,\,50$\,Mpc. For extinction, we consider $A_{\mathrm{V}}=0$ and 1\,mag and use a standard extinction law, as described in \cite{Cardelli-1989}, with $R_V=3.1$. For host-galaxy background, we consider an $R$-band surface brightness of $\mu_R=+\infty$ (i.e., no background) and 20\,mag\,arcsec$^{-2}$; we calculate the surface brightnesses in the other filters by assuming an SED of a typical Sb spiral galaxy from the \textsc{swire} library \cite{Polletta-2007}, with which we re-derive the detection limits of \textit{Euclid}/CSST.

As previously mentioned, at $d=10$\,Mpc without extinction or host-galaxy background the lowest-mass RSG progenitor with $M_{\mathrm{ini}}=8.0\,M_\odot$ can be detected in all \textit{Euclid}/CSST filters except the very blue $NUV$ and $u$ bands. At 30\,Mpc, an 8-$M_\odot$ progenitor is still detectable in the $r,\,i,\,z,\,y,\,I$ bands and its brightness falls below the detection limits of all other filters. Now the $g$ band can detect down to $M_{\mathrm{ini}}=11.8\,M_\odot$. At 50\,Mpc, an 8-$M_\odot$ progenitor is no longer detectable, and $9.0\,M_\odot$ is the lowest mass of RSG progenitors that can be detected in at least one filter (the $i$ band) while detection in the $z$ band requires a slightly higher initial mass of $9.1\,M_\odot$.

At $d=10$\,Mpc, the effect of extinction is not very obvious since at such a close distance the RSG progenitors are very bright even with a higher extinction. An 8-$M_\odot$ progenitor can still be detected in most of the filters. At larger distances, the effect of extinction becomes pronounced for the bluer filters. For instance, at 30\,Mpc, the $g$ band can only detect down to 21.8\,$M_\odot$ with $A_{\mathrm{V}}=1$\,mag compared with 11.8\,$M_\odot$ without extinction. Extinction hardly affects the redder, in particular the NIR filters. For instance, at $d=50\,\mathrm{Mpc}$ and $A_{\mathrm{V}}=1\,\mathrm{mag}$, the lowest mass to be detected in the $J$ and $H$ bands only slightly increase to 19.5 and 18.0\,$M_\odot$ from 17.1 and 16.7\,$M_\odot$ ($A_{\mathrm{V}} = 0$), respectively, due to their low extinction coefficients ($A_{\mathrm{J}}/A_{\mathrm{V}} \approx 0.25$, $A_{\mathrm{H}}/A_{\mathrm{V}} \approx 0.17$).

Increasing the host-galaxy background degrades the detection limits of point sources by \textit{Euclid}/CSST. At $d=10$\,Mpc, the effect of such a surface brightness is not obvious since at such a close distance the RSG progenitors are very bright. At $d=30$\,Mpc, the lowest detectable masses in the $g, r,\,i,\,z,\,y$ bands are 17.4, 10.1, 8.4, 8.5, and 9.2\,$M_\odot$, indicating that these bands still offer strong detection capabilities. In the $Y$, $J$ and $H$ bands, however, the lowest mass to be detected increases significantly from 10.9, 9.4 and 9.3\,$M_\odot$ (no host-galaxy background) to 24.3, 21.7 and 22.6\,$M_\odot$, respectively, indicating that the effect of host-galaxy background becomes pronounced in the NIR filters. At $d=50$\,Mpc, even a 30-$M_\odot$ star becomes undetectable in the $Y$, $J$ and $H$ bands. This wavelength-dependent effect arises primarily because, in spiral galaxies, redder bands exhibit significantly higher surface brightness than bluer bands due to the presence of a dust-rich stellar disk.

In summary, the \textit{Euclid} and CSST filters are very useful in detecting the canonical Type~II-P progenitors (i.e., single-star RSGs) except the very blue $NUV$ and $u$ bands. At a distance of 10\,Mpc, an 8-$M_\odot$ progenitor can be detected in all other filters in case of no extinction or host-galaxy background. With higher distance, extinction and/or host-galaxy background (e.g., scenarios involving binary evolution, enhanced circumstellar mass loss, or dense dusty envelopes), the detection capability for RSG progenitors will be partially degraded, yet detection potential remains.

Notably, in crowded stellar fields, particularly near the centers of galaxies or in dense star-forming regions, source confusion and background contamination can hinder the identification and photometric measurement of individual RSG progenitors. Overlapping on point-spread functions (PSFs) makes it difficult to disentangle the progenitor’s signal from neighboring stars or diffuse emission. Thus, while the angular resolution and the filters themselves are well-suited for RSG detection, practical detectability also depends critically on the local stellar density and the effectiveness of photometric techniques in crowded environments. In this work, we do not analyze the effect of confusion with other resolved objects, which requires detailed image simulations and is beyond the scope of this work. As a result, our estimate of the detection capability is overrated.

\subsection{Detection Rates}\label{rate}

\begin{figure}[H]
    \centering
    \includegraphics[width=\columnwidth]{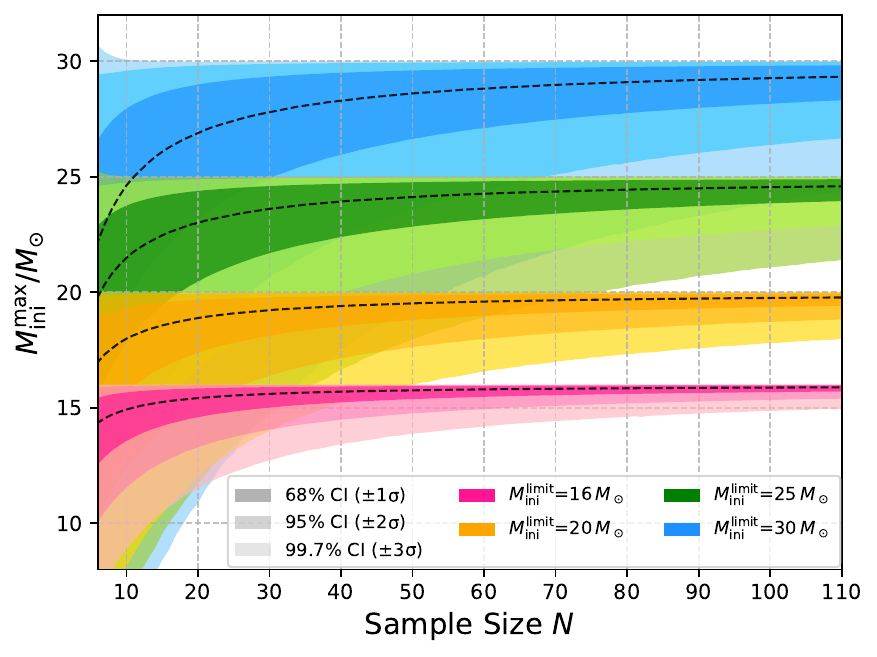}
    \caption{Expected maximum initial masses $M_\mathrm{ini}^\mathrm{max}$ in SN II-P progenitor samples of varying sizes, derived with different assumed upper limits on the initial masses $M_\mathrm{ini}^\mathrm{limit}$ of RSGs. The pink, yellow, green, and blue shaded regions represent the uncertainty ranges for samples with $M_\mathrm{ini}^\mathrm{limit}=16,\,20,\,25$ and $30\,M_{\odot}$, respectively. Within each colored region, dashed lines indicate the median values for each sample group, and the shading intensity varies from dark to light to represent different confidence intervals (CIs): dark, medium, and light shadings reflect the 68\%, 95\%, and 99.7\% CIs, respectively, corresponding to $\pm 1\sigma$/$\pm 2\sigma$/$\pm 3\sigma$ deviations from the median.}
    \label{fig:sample}
\end{figure}

After assessing the detection capabilities of SN progenitors using the \textit{Euclid} and CSST surveys, a key remaining question pertains to the detectable samples: specifically, the annual detection rate of progenitors and the scientific insights that can be derived from these samples. To estimate the annual detection rates for Type~II-P SN progenitors with different assumed upper mass limits in the \textit{Euclid} and CSST surveys, we employ Monte Carlo simulations to generate random stellar samples within the surveyed galaxies and evaluate their detectability.

Initial stellar masses are first randomly generated with a power-law index of $-2.35$ based on the initial mass function (IMF) introduced by \cite{Salpeter-1955}, followed by a selection process that retains only those within a specific initial mass range (such as 8--16\,$M_{\odot}$ or 8--25\,$M_{\odot}$, etc.) as Type~II-P SN progenitors. These selected progenitor stars are then assigned to galaxies within the survey area of \textit{Euclid} and CSST using the galaxies' star formation rate (SFR) as weighting factors. The galaxy catalog used in this study is from the \textsc{z0mgs} project \cite{z0mgs-Leroy-2019, z0mgs-2019}. SFRs of galaxies are also from the \textsc{z0mgs} project while Galactic extinction values are calculated using \textsc{dustmaps} package \cite{dustmaps-Green-2018} based on locations from \textsc{z0mgs} and the two-dimensional extinction maps from \cite{sfd-Schlegel-1998, sfd-Schlafly-2011}.

To simulate the host-galaxy background brightness around generated SN progenitors in the survey bands of the \textit{Euclid} and CSST, it is essential to model the radial positions of progenitor stars within a galaxy and the surface brightness profiles of galaxies for the corresponding filters. 
In an exponential disk model, the surface density of star formation rate typically follows $\Sigma_{\mathrm{SFR}}(r) \propto \exp(-r/r_0)$, where $r_0$ is the scale length. The total star formation rate in an annulus of radius $r$ and width $dr$, $2\pi r \Sigma_{\mathrm{SFR}}(r) \, dr$, is proportional to $r \exp(-r/r_0) \, dr$. When normalized, this radial distribution defines a probability density function (PDF) for the radius at which a randomly selected star-forming event occurs. Introducing the dimensionless variable $x = r/r_0$, the resulting PDF is $P(x) \propto x e^{-x}$ for $x \geq 0$, which corresponds to a Gamma distribution with shape parameter $k = 2$ and scale parameter $\theta = 1$. Thus, the stellar formation radius (in units of the scale length) follows a Gamma(2, 1) distribution. Based on this distribution, we simulate the radial positions of progenitor stars within the galaxies.

Next, to determine the background brightness levels at corresponding radius in the \textit{Euclid} and CSST filters, observed surface brightness profiles and SEDs for all galaxies within the survey footprints are required. Despite the variation of surface brightness profiles and SEDs among galaxies, there remains practical difficulty of obtaining them for large galaxy samples. Therefore, we adopt an observed surface brightness profile from a typical spiral galaxy with a matching SED template as a rough but pragmatic evaluation. Specifically, we utilize the surface brightness profile of NGC~7819 (UGC~26), a face-on Sc-type galaxy \cite{Kalinova-2021}, as provided by \cite{Gilhuly-2018} from the Calar Alto Legacy Integral Field Spectroscopy Area (CALIFA) survey's third data release (DR3) \cite{CALIFA-Sanchez-2012, CALIFA-Walcher-2014, CALIFA-Sanchez-2016}. Subsequently, we apply a type Sc galaxy SED template from the \textsc{swire} library \cite{Polletta-2007} to transform the observed SDSS $i$-band surface brightness profile into the respective survey bands of \textit{Euclid} and CSST. We have also conducted a test by changing surface brightness profile to an Sa-type galaxy NGC~5614 and received a closely similar result of detection rate.

For each simulated progenitor, we adopt the effective temperature ($T_\mathrm{eff}$) and bolometric luminosity ($L_\mathrm{bol}$) at its 
final evolutionary stage from the \textsc{parsec} stellar evolution models , using the corresponding initial mass. Based on these parameters, progenitor's theoretical magnitudes are derived by interpolating from the \textsc{marcs} model atmospheres \cite{Gustafsson-2008}, assuming a microturbulent velocity of $5\,\mathrm{km\,s^{-1}}$, surface gravity $\log(g)=0$ and solar metallicity $[\mathrm{Fe/H}]=0$. We then determine the detection limiting magnitudes with the background brightness and the Galactic extinction for the position of each simulated progenitor and compare the limits with the progenitor's theoretical magnitudes. Based on the above analysis, we predict that, upon completion of the \textit{Euclid} and CSST surveys, $\lesssim13$ (or $24$) Type~II-P SN progenitors with the initial mass range of 8--16 (or 8--25)\,$M_{\odot}$ will be detected annually from the archival survey images with SNR $\geq5$.

The above estimate assumes neither internal extinction within the host galaxy nor detection limits degradation due to source crowding. It is, however, very challenging to accurately predict their values and effects for the simulated progenitors. As a rough estimate, we carry out two more simulations. In one simulation, we assume all progenitors suffer from a host extinction reddening of $E(B-V)=0.1$\,mag (which is a typical value) and the annual progenitor detection rate decreases to 11 (or 22) for the initial mass range of 8--16 (or 8--25)\,$M_{\odot}$. In the other simulation, we conservatively assume that the detection limits become shallower by 1\,mag due to source crowding and the rate decreases to 5 (or 9) for the initial mass range of 8--16 (or 8--25)\,$M_{\odot}$. Therefore, internal extinction and source crowding will reduce the number of detected progenitors.

\subsubsection{Significance of the RSG problem}
Due to the limited size of observed Type~II-P SN progenitor samples, it remains unclear whether the RSG problem reflects a true physical upper mass limit for progenitor stars or is instead an artifact of small-number statistics and observational biases. It was emphasized in \cite{Davies-2020} that the statistical significance of the RSG problem is between 1$\sigma$ and 2$\sigma$ as it could be consistent with stochastic sampling from a continuous IMF rather than a physical boundary. In the future, with a much larger sample of RSG progenitors detected by \textit{Euclid} and CSST surveys, a key question that can be answered is the statistical significance of the upper limits on the initial masses $M_\mathrm{ini}^\mathrm{limit}$ of Type~II-P SN progenitors.

To clarify this topic, we perform statistical simulations to show expected maximum initial masses $M_\mathrm{ini}^\mathrm{max}$ in progenitor samples of varying sizes. We generated initial masses of stars based on a IMF introduced by \cite{Salpeter-1955} and then selected progenitor samples according to different assumed upper limits on the initial masses, $M_\mathrm{ini}^\mathrm{limit}$, of SN II-P progenitors. From these samples, we randomly drew subsamples of varying sizes and recorded their maximum initial masses. After performing a large number of repeated samplings, we obtained the relation between the maximum initial mass and the sample size for different upper mass limits, as shown in Figure~\ref{fig:sample}.

For a sample size of $N\sim25$ (i.e., the currently observed SN II-P progenitor sample), the maximum initial masses corresponding to different assumed upper mass limits are not significantly distinguishable. However, with a sample size of $N\gtrsim35$, we can statistically distinguish between $M_\mathrm{ini}^\mathrm{limit}=16$ and $25\,M_{\odot}$ at 3$\sigma$ significance. Assuming an expected practical annual detection rate of approximately 10 SN II-P progenitors, such a sample would require about 3\,years to accumulate.

\section{Estimating Parameters of RSG Progenitors}\label{parameters}

\begin{figure}[H]
    \centering
    \includegraphics[width=\columnwidth]{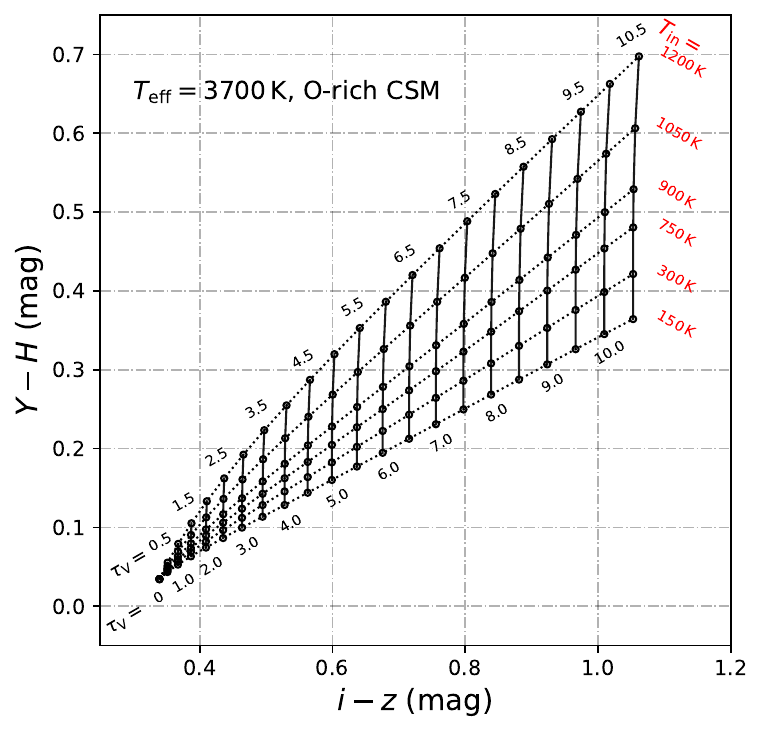}
    \caption{$Y-H$ versus $i-z$ color-color diagram for the RSG of $T_{\rm eff}=3700\,\mathrm{K}$ with O-rich circumstellar dust of different $T_\mathrm{in}$ and $\tau_\mathrm{V}$.}
    \label{fig:mock_ccd}
\end{figure}

\begin{figure*}
    \centering
    \includegraphics[width=\textwidth]{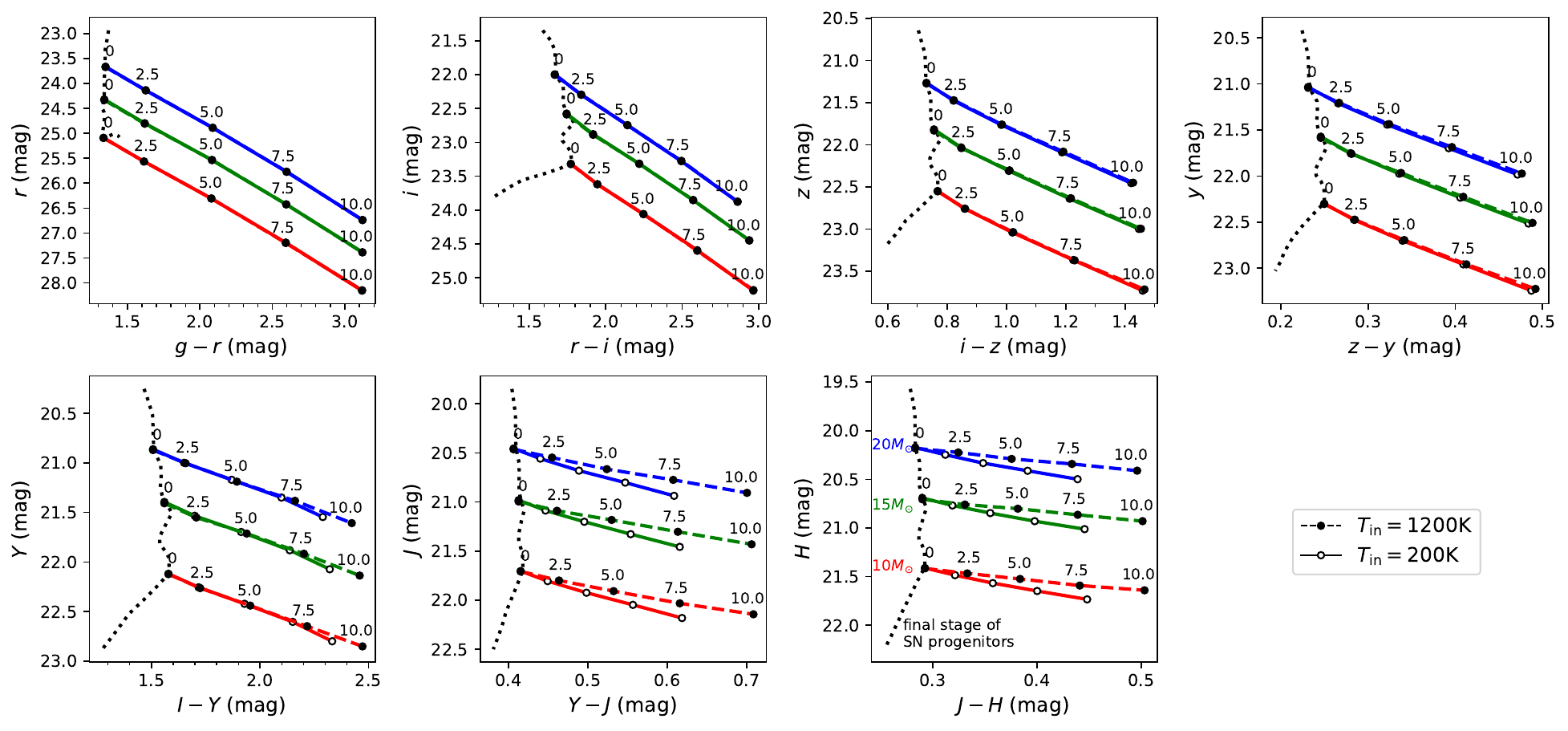}
    \caption{Color-magnitude diagrams for RSGs with initial masses $M_\mathrm{ini}=10,\,15,\,20\,M_{\odot}$ embedded in O-rich circumstellar dust characterized by various inner temperature $T_\mathrm{in}$ and V-band optical depth $\tau_\mathrm{V}$. Black dotted lines in the left portion of each panel correspond to the final stage of SN progenitors with different initial masses assuming no extinction. Colors denote the initial mass of the progenitor: red for $10\,M_{\odot}$, green for $15\,M_{\odot}$, and blue for $20\,M_{\odot}$. Solid and dashed lines represent $T_\mathrm{in} = 200\,\mathrm{K}$ and $T_\mathrm{in} = 1200\,\mathrm{K}$, respectively, with data points labeled with their corresponding $\tau_\mathrm{V}$ values.}
    \label{fig:mock_cmd}
\end{figure*}

\begin{figure*}
    \centering
    \includegraphics[width=\textwidth]{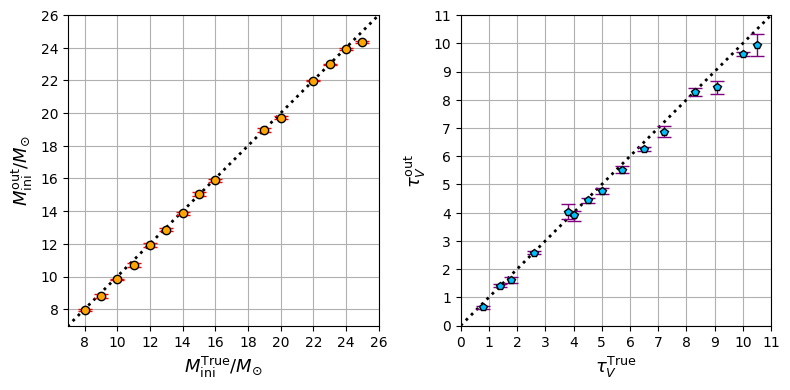}
    \caption{Inferred parameters from MCMC fitting in comparison with the input values. \textit{Left}: $M_\mathrm{ini}$, \textit{Right}: $\tau_\mathrm{V}$.}
    \label{fig:mock_compare}
\end{figure*}

After detecting SN progenitors, it is important to derive their fundamental stellar parameters. Such parameters include the effective temperature, bolometric luminosity and the initial mass, etc.
However, the existence and uncertainty of circumstellar extinction makes this difficult. Circumstellar dust plays three different roles in affecting the SED of the SN progenitor: (1) it can attenuate the stellar radiation along the line of sight via absorption and scattering (\textit{attenuated radiation}); (2) it may scatter the stellar radiation along other directions into the line of sight, partially compensating the loss of light (\textit{scattered radiation}); (3) it can also generate its own emission, particularly at infrared wavelengths, which is sensitive to the dust temperature (\textit{dust emission}).

To evaluate the effect of circumstellar extinction on the SED of RSG progenitors as observed by \textit{Euclid} and CSST, we synthesize the SEDs of RSGs with an initial mass of $M_\mathrm{ini}=15\,M_{\odot}$, surrounded by spherically distributed O-rich dust envelopes.
For the intrinsic radiation, the effective temperature $T_\mathrm{eff}=3374\,\mathrm{K}$ and bolometric luminosity, $\log(L_\mathrm{bol}/L_{\odot})=5.087$, which characterize the final evolutionary stage of a 15-$M_{\odot}$ progenitor, are adopted from the \textsc{parsec} stellar evolution models. We then obtain the intrinsic SED by interpolating from the \textsc{marcs} model atmospheres \cite{Gustafsson-2008}, assuming a microturbulent velocity of $5\,\mathrm{km\,s^{-1}}$, surface gravity $\log(g)=0$ and solar metallicity $[\mathrm{Fe/H}]=0$.

For the circumstellar dust, we assume a $\rho \propto r^{-2}$ wind-like radial density profile, a default relative thickness that is 1000 times the inner radius and a standard grain size distribution proposed by \cite{Mathis-1977}, while the dust temperature at the inner boundary ($T_\mathrm{in}$) and the V-band optical depth of the dust envelope ($\tau_\mathrm{V}$) are selected as the two free parameters to study their effects. The radiation transfer through the circumstellar dust is solved with the \textsc{dusty} \cite{dusty-Ivezic-1997} code and the output SEDs are showed in Figure~\ref{fig:sed_m_sun_15}.

From Figure~\ref{fig:sed_m_sun_15}, it is clear that scattered radiation dominates the total emerging radiation from the $NUV$ to $y$ bands, attenuated radiation dominates in the $J$ and $H$ bands at longer wavelengths, while in the $Y$ band, the scattered radiation and attenuated radiation have comparable contributions to the total emerging radiation. Both the scattered radiation and attenuated radiation are sensitive to optical depth $\tau_{\rm V}$ and do not depend on dust temperature $T_{\rm in}$. In comparison, dust emission depends on both $\tau_{V}$ and $T_{\rm in}$; however, the contribution of dust emission is very small even in the longest-wavelength $H$ band ($\sim$10\% for $T_{\rm in}=1200$\,K, down to close to zero for $T_{\rm in}=200$\,K). As a result, the total emerging radiation is primarily dependent on the optical depth $\tau_{\rm V}$ and is very insensitive to dust temperature $T_{\rm in}$ in the \textit{Euclid} and CSST filters.

This effect is also obvious in the color-color diagram for an RSG of $T_\mathrm{eff}=3700\,$K with O-rich circumstellar dust (see Figure~\ref{fig:mock_ccd}). 
As $\tau_\mathrm{V}$ increases, both the $Y-H$ and $i-z$ colors become redder, reflecting the reddening effect. The $i-z$ color remains nearly constant across different values of $T_\mathrm{in}$, while the $Y-H$ color turns redder with increasing $T_\mathrm{in}$. This is consistent with the above analysis: dust emission, which depends on temperature, has a weak contribution only in the $H$ band while in the other filters dust emission is negligible. However, even at a large $\tau_\mathrm{V}$ of 10, where the color differences due to different $T_\mathrm{in}$ values are relatively pronounced, the variation in $Y-H$ color is only about 0.4\,mag for $T_\mathrm{in}$ ranging from 150\,K to 1200\,K (which is required to be lower than 1500\,K such that the dust can survive in the envelope; \cite{Pozzo-2004, Sarangi-2018}). This difference is not very significant, considering that SN progenitors are often detected just above the detection limit and with large photometric uncertainties \cite[e.g.,][]{Niu-2023ixf-2023}. This indicates that the circumstellar dust’s $T_\mathrm{in}$ has negligible impact on the progenitor’s color.

In summary, in the presence of circumstellar dust, $\tau_\mathrm{V}$ controls the SED of the total emerging radiation of the SN progenitor while $T_\mathrm{in}$ has little effect for the \textit{Euclid} and CSST filters. Based on this, we propose a new method: taking advantage of the extensive 11 filters of the \textit{Euclid} and CSST surveys, one can fit $M_{\mathrm{ini}}$ and $\tau_\mathrm{V}$ simultaneously by fixing $T_{\rm in}$ to a typical value and tracing the observed SED back to the intrinsic one. The color-magnitude diagram (see Figure~\ref{fig:mock_cmd}) demonstrates how this approach enables the recovery of the progenitor's parameters through comparison with outputs from stellar evolution models such as \textsc{parsec}.

To validate the robustness of our new method, we conduct mock data tests by randomly generating a set of initial masses ($M_{\mathrm{ini}}$), dust temperatures at the inner boundary ($T_\mathrm{in}$) and V-band optical depths ($\tau_{\mathrm V}$) for RSGs with circumstellar dust. We simulate their multi-band photometry based on the \textsc{marcs} model atmospheres \cite{Gustafsson-2008} to generate synthetic magnitudes and add photometric uncertainties. 
For these mock observational data, we then try to recover their $M_{\rm ini}$ and $\tau_{\rm V}$ by multi-band SED fitting while fixing $T_{\rm in}$ to a typical value. Here, we set $T_{\rm in}=1000\,{\rm K}$ and generate model SEDs for different $M_{\rm ini}$ and $\tau_{\rm V}$ with the same method as above. We then use a Bayesian + Markov Chain Monte Carlo (MCMC) method to fit model SEDs to the mock observational data and search for the best-fitting values of $M_{\rm ini}$ and $\tau_{\rm V}$. Flat priors are used for both parameters, and the posterior probability distributions give their most probable values as well as the credible intervals. As shown in Figure~\ref{fig:mock_compare}, the inferred parameters agree remarkably well with the input values. This consistency confirms that multi-band SED fitting alone is sufficient to accurately constrain the intrinsic properties of RSG progenitors. The success can be attributed to the fact that RSGs exhibit similar intrinsic effective temperatures and photometric colors in their pre-SN stages and that their SEDs are critically dependent on $M_{\mathrm{ini}}$ and $\tau_{\mathrm V}$ and is almost independent of $T_\mathrm{in}$. This result strongly validates the reliability and precision of our proposed methodology.

Some recently identified SN progenitors have been inferred to be extremely reddened by circumstellar dust. For example, the progenitor of SN~2023ixf, located at a distance of approximately 7\,Mpc, is inferred to be surrounded by circumstellar dust with an optical depth of $\tau_{\mathrm V}\approx5$--13 \cite{Kilpatrick-2023ixf-2023, Niu-2023ixf-2023, Neustadt-2023ixf-2024, Qin-2023ixf-2024, VanDyk-2023ixf-2024, Xiang-2023ixf-2024}. Similarly, the progenitor of SN~2025pht, at a distance of about 12\,Mpc, has been reported to be surrounded by dust with $\tau_{\mathrm V}\approx5$--14 \cite{Kilpatrick-2025pht-2025, VanDyk-2025pht-2026}. The lower values of $\tau_{\mathrm V}$ are inferred under the assumption of C-rich (graphite) dust, whereas the higher values correspond to O-rich (silicate) dust. Graphite exhibits significantly higher absorption efficiency in the NIR, allowing the observed SED to be reproduced with a relatively small optical depth. In contrast, silicate dust is much more transparent at NIR wavelengths and therefore requires a substantially larger optical depth to produce a similar SED.

Our simulations indicate that such highly reddened progenitors should remain detectable and characterizable at distances of ~10\,Mpc. As discussed in the first paragraph of this section, scattered radiation partially compensates for the loss of stellar light, reducing the net attenuation in the total observed radiation. Figure~\ref{fig:sed_m_sun_15} shows that the flux in the V band ($\sim5500\,\text{Å}$) decreases by a factor of approximately 70 between the $\tau_{\mathrm V}$=0 and $\tau_{\mathrm V}$=10 models, corresponding to a dimming of about 4.5\,mag. Moreover, the NIR flux is much less affected by increasing dust optical depth. Therefore, at distances of around 10\,Mpc, identifying such highly reddened progenitors should remain feasible by \textit{Euclid} and CSST, either in deep optical filters or through the NIR filters.
In our former analysis, we explored models with optical depths up to 10.5. Additional tests indicate that extending the models to even larger optical depths does not affect the accuracy of the mass-inference procedure. This is because the emission from the circumstellar dust remains weak, making 
the effects of increasing optical depths on the \textit{Euclid} and CSST filters relatively straightforward to model.

\section{Summary and Discussion}\label{conclusion}
In this work, we evaluate the capability of \textit{Euclid} and CSST surveys in detecting Type~II-P SN progenitors and explore how to robustly estimate their intrinsic parameters in the presence of circumstellar dust. We find that the optical and near-infrared filters of the \textit{Euclid} and CSST are highly effective for detecting RSG progenitors of Type~II-P SNe while the very blue $NUV$ and $u$ bands are less useful. At a distance of 10\,Mpc, an 8-$M_\odot$ progenitor is detectable in all other filters with no extinction or host-galaxy background. However, the detectability diminishes with increasing distance, as well as in the presence of extinction and/or host-galaxy background. 

Under optimistic assumptions, we predict that the archival data from the completed \textit{Euclid} and CSST surveys will yield $\lesssim13$ (or $24$) detections of Type~II-P SN progenitors per year with the initial mass range of 8--16 (or 8--25)\,$M_{\odot}$ at SNR $\geq5$. This is a very optimistic estimate since we have not considered internal extinction within the host galaxy or confusion with other resolved sources. Furthermore, a progenitor sample of size $N\gtrsim35$ would statistically distinguish between upper mass limits of $M_\mathrm{ini}^\mathrm{limit}=16$ and $25\,M_{\odot}$ with a significance of 3$\sigma$, which is expected to be accumulated within about 3\,years.

Our analysis further demonstrates that the V-band optical depth of circumstellar dust ($\tau_\mathrm{V}$), is the dominant factor governing the SED of the total emerging radiation of SN progenitor in the \textit{Euclid} and CSST filters, whereas the dust temperature at the inner radius ($T_\mathrm{in}$) exerts little influence on the SED in these filters. Given that RSGs exhibit remarkably uniform effective temperatures and photometric colors during their pre-SN phases, their SEDs are primarily sensitive to $M_\mathrm{ini}$ and $\tau_\mathrm{V}$ and is almost independent of $T_\mathrm{in}$. This finding underscores the robustness of multi-band SED fitting for constraining the intrinsic properties of RSG progenitors.

Moreover, the pulsational variability of RSGs represents a critical yet often overlooked systematic uncertainty in the derivation of progenitor parameters for Type~II-P SNe. These oscillations, which are commonly detected in nearby Type~II-P SN progenitors such as SN~2023ixf \cite{Niu-2023ixf-2023, Xiang-2023ixf-2024} and SN~2024ggi \cite{Xiang-2024ggi-2024}, induce time-dependent variations in luminosity, effective temperature, and radius, thereby biasing estimates of progenitor parameters derived from pre-explosion imaging. The \textit{Euclid} and CSST surveys are independent projects that, despite having overlapping survey footprints, operate with non-simultaneous observing schedules, so the variability will increase the complexity of parameter estimation analysis. However, the multi-epoch nature of their survey strategies (comprising 4 epochs in $NUV$ and $y$ bands and 2 epochs in the other filters for CSST; 3-4 epochs for \textit{Euclid}) may help constrain RSG pulsations and thus better characterize the final evolutionary stage of massive stars.

In addition to \textit{Euclid} and CSST, other telescopes will also play important roles in the search for SN progenitors, such as HST \cite[e.g.,][]{Maund-2005}, JWST \cite[e.g.,][]{Kilpatrick-2025pht-2025}, LSST \cite{Strotjohann-2024} and \textit{Roman} Space Telescope. Specifically, \textit{Euclid} and CSST offer better spatial resolution and Northern Hemisphere coverage compared to LSST, significantly broader sky coverage and wavelength range than JWST, and a wider survey area than \textit{Roman} Space Telescope. However, rather than viewing these missions as competing entities, it is crucial to recognize their synergistic potential. The high-resolution imaging from CSST and Euclid can be effectively combined with the deep, multi-epoch time-domain data from LSST and the sensitive infrared data from JWST and \textit{Roman}. This wealth of data will enable the detection of SN progenitors over a wide wavelength range, at high spatial resolutions, down to deep detection limits, and across a large number of nearby galaxies, providing new insights into the evolution of massive stars and the origin of SNe.

\Acknowledgements{The authors acknowledge the help from Prof. Mark Cropper, Dr. Mischa Schirmer, Dr. Erik Romelli, Dr. Paola Maria Battaglia, Dr. Chiara Sirignano and Dr. Eduardo Medinaceli for discussions and suggestions regarding the exposure time calculators of VIS and NISP of \textit{Euclid}. NCS is funded by the China Manned Space Program No.CMS-CSST-2025-A14, the National Natural Science Foundation of China (NSFC) Grants No.12303051 and No.12261141690 and the Strategic Priority Research Program of the Chinese Academy of Sciences Grant No.XDB0550300. TMZ is funded by the NSFC Grant No.12233008. We acknowledge the use of the following software packages, programs, datasets, and model libraries: \textsc{astropy}, \textsc{califa}, \textsc{dustmaps}, \textsc{dusty}, \textsc{marcs}, \textsc{parsec}, \textsc{pysynphot}, \textsc{swire} Template Library, and \textsc{z0mgs}.}

\InterestConflict{The authors declare that they have no conflict of interest.}
\printAcknowledgements
\printInterestConflict

\bibliographystyle{scpma}
\bibliography{reference}




\end{multicols}
\end{document}